\documentclass[twocolumn]{aastex62}
\usepackage{verbatim}
\usepackage{graphicx}
\usepackage{multirow}

\newcommand{\methanol}{CH$_3$OH}

\newcommand{\lsun}{$L_\odot$}
\newcommand{\msun}{$M_\odot$}

\newcommand{\mjb}{mJy~beam$^{-1}$}

\newcommand{\kms}{\mbox{km~s$^{-1}$}}

\newcommand{\eupper}{$E_{\rm upper}/k$}

\newcommand{\ngci}{NGC6334I}

\newcommand{\mum}{$\mu$m}

\newcommand{\gtfe}{G358.93$-$0.03}

\newcommand{\vt}{${\rm v}_t$}

\newcommand{\vlsr}{$v_{\rm LSR}$}

\submitjournal{ApJ Letters}
\received{2019 May 31}
\revised{2019 June 30}
\accepted{2019 July 04}

\shorttitle{Discovery of 14 new methanol maser lines in \gtfe}
\shortauthors{Brogan et al.}

\begin{document}

\title{Sub-arcsecond (sub)millimeter imaging of the massive protocluster \gtfe:\\ Discovery of 14 new methanol maser lines associated with a hot core}


\author{C. L. Brogan}
\affiliation{National Radio Astronomy Observatory (NRAO), 520 Edgemont Rd, Charlottesville, VA 22903, USA} 

\author{T. R. Hunter}
\affiliation{National Radio Astronomy Observatory (NRAO), 520 Edgemont Rd, Charlottesville, VA 22903, USA}

\author{A. P. M. Towner}
\altaffiliation{A.P.M.T. is a Grote Reber Doctoral Fellow at NRAO} 
\affiliation{National Radio Astronomy Observatory (NRAO), 520 Edgemont Rd, Charlottesville, VA 22903, USA}
\affiliation{Department of Astronomy, University of Virginia, P.O. Box 3818}

\author{B. A. McGuire}
\altaffiliation{B.A.M. is a Hubble Fellow of NRAO} 
\affiliation{National Radio Astronomy Observatory (NRAO), 520 Edgemont Rd, Charlottesville, VA 22903, USA}
\affiliation{Center for Astrophysics $\mid$ Harvard~\&~Smithsonian, Cambridge, MA 02138, USA}

\author{G. C. MacLeod}
\affiliation{Hartebeesthoek Radio Astronomy Observatory, PO Box 443, Krugersdorp 1740, South Africa}
\affiliation{The University of Western Ontario, 1151 Richmond Street, London, ON N6A 3K7, Canada}

\author{M. A. Gurwell}
\affiliation{Center for Astrophysics $\mid$ Harvard~\&~Smithsonian, Cambridge, MA 02138, USA}

\author{C. J. Cyganowski}
\affiliation{SUPA, School of Physics and Astronomy, University of St. Andrews, North Haugh, St. Andrews KY16 9SS, UK}

\author{J. Brand}
\affiliation{INAF-Istituto di Radioastronomia, via P. Gobetti 101, 40129 Bologna, Italy}
\affiliation{Italian ALMA Regional Centre, via P. Gobetti 101, 40129 Bologna, Italy}

\author{R. A. Burns}
\affiliation{Mizusawa VLBI Observatory, National Astronomical Observatory of Japan, 2-21-1 Osawa, Mitaka, Tokyo 181-8588, Japan}
\affiliation{Korea Astronomy and Space Science Institute, 776 Daedeokdae-ro, Yuseong-gu, Daejeon 34055, Republic of Korea}

\author{A. Caratti o Garatti}
\affiliation{Dublin Institute for Advanced Studies, 31 Fitzwilliam Place, D02 XF86, Dublin, Ireland}

\author{X. Chen}
\affiliation{Center for Astrophysics, GuangZhou University, Guangzhou 510006, China}

\author{J. O. Chibueze}
\affiliation{Space Research Unit, Physics Department, North-West University, Potchefstroom, 2520, South Africa}
\affiliation{Department of Physics and Astronomy, University of Nigeria, Carver Building, 1 University Road, Nsukka, 410001, Nigeria}

\author{N. Hirano}
\affiliation{Institute of Astronomy and Astrophysics, Academia Sinica, 11F of Astronomy-Mathematics Building, AS/NTU, No.1, Sec. 4, Roosevelt Rd, Taipei 10617, Taiwan, R.O.C.}

\author{T. Hirota}
\affiliation{Mizusawa VLBI Observatory, National Astronomical Observatory of Japan, 2-21-1 Osawa, Mitaka, Tokyo 181-8588, Japan}
\affiliation{Department of Astronomical Sciences, SOKENDAI (The Graduate University for Advanced Studies), Osawa 2-21-1, Mitaka-shi, Tokyo 181-8588, Japan}

\author{K.-T. Kim}
\affiliation{Korea Astronomy and Space Science Institute, 776 Daedeokdae-ro, Yuseong-gu, Daejeon 34055, Republic of Korea}

\author{B. H. Kramer}
\affiliation{Max-Planck-Institut f\"{u}r Radioastronomie, Auf dem H\"{u}gel 69, D-53121 Bonn, Germany}
\affiliation{National Astronomical Research Institute of Thailand, 260 Moo 4, T. Donkaew, Amphur Maerim, Chiang Mai, 50180, Thailand}

\author{H. Linz}
\affiliation{Max-Planck-Institut f\"{u}r Astronomie, K\"onigstuhl 17, 69117 Heidelberg, Germany}

\author{K. M. Menten}
\affiliation{Max-Planck-Institut f\"{u}r Radioastronomie, Auf dem H\"{u}gel 69, D-53121 Bonn, Germany}

\author{A. Remijan}
\affiliation{National Radio Astronomy Observatory (NRAO), 520 Edgemont Rd, Charlottesville, VA 22903, USA}

\author{A. Sanna}
\affiliation{Max-Planck-Institut f\"{u}r Radioastronomie, Auf dem H\"{u}gel 69, D-53121 Bonn, Germany}

\author{A. M. Sobolev}
\affiliation{Ural Federal University, 19 Mira street, 620002 Ekaterinburg, Russia }

\author{T. K. Sridharan}
\affiliation{Center for Astrophysics $\mid$ Harvard~\&~Smithsonian, Cambridge, MA 02138, USA}

\author{B. Stecklum}
\affiliation{Th\"{u}ringer Landessternwarte, Sternwarte 5, 07778 Tautenburg, Germany}

\author{K. Sugiyama}
\affiliation{National Astronomical Observatory of Japan, Osawa 2-21-1, Mitaka-shi, Tokyo 181-8588, Japan}

\author{G. Surcis}
\affiliation{INAF-Osservatorio Astronomico di Cagliari, Via della Scienza 5, I-09047, Selargius, Italy}

\author{J. Van der Walt}
\affiliation{Centre for Space Research, North-West University, Potchefstroom, South Africa}

\author{A. E. Volvach}
\affiliation{Radio Astronomy Laboratory of Crimean Astrophysical Observatory, Katsively, RT-22 Crimea}
    \affiliation{Institute of Applied Astronomy, Russian Academy of Sciences, St. Petersburg, Russia}

\author{L. N. Volvach}
\affiliation{Radio Astronomy Laboratory of Crimean Astrophysical Observatory, Katsively, RT-22 Crimea}

\correspondingauthor{C. L. Brogan}
\email{cbrogan@nrao.edu}

\begin{abstract}
We present (sub)millimeter imaging at $0\farcs5$ resolution of the massive star-forming region \gtfe\/ acquired in multiple epochs at 2 and 3 months following the recent flaring of its 6.7~GHz \methanol\/ maser emission. Using SMA and ALMA, we have discovered 14 new Class\,II \methanol\/ maser lines ranging in frequency from 199 to 361\,GHz, which originate mostly from \vt=1 torsionally-excited transitions and include one \vt=2 transition.  The latter detection provides the first observational evidence that Class\,II maser pumping involves levels in the \vt=2 state.
The masers are associated with the brightest continuum source (MM1), which hosts a line-rich hot core.  The masers present a consistent curvilinear spatial velocity pattern that wraps around MM1, suggestive of a coherent physical structure 1200\,au in extent.  In contrast, the thermal lines exhibit a linear pattern that crosses MM1 but at progressive position angles that appear to be a function of either increasing temperature or decreasing optical depth. The maser spectral profiles evolved significantly over one month, and the intensities dropped by factors of 3.0 to 7.2, with the \vt=2 line showing the largest change.  A small area of maser emission from only the highest excitation lines closest to MM1 has disappeared. There are seven additional dust continuum sources in the protocluster, including another hot core (MM3).  We do not find evidence for a significant change in (sub)millimeter continuum emission from any of the sources during the one month interval, and the total protocluster emission remains comparable to prior single dish measurements. 
\end{abstract}


\section{Introduction}

\begin{deluxetable*}{ccccccccc}
\tabletypesize{\scriptsize}
\tablecolumns{8}
\setlength{\tabcolsep}{0.5mm}
\tablecaption{Observations summary \label{tab:observations}}
\tablehead{\colhead{Band(s)} & \colhead{Date}  &  \colhead{Time on} & \colhead{PWV\tablenotemark{a}} & \colhead{uv range\tablenotemark{b}} & \colhead{Spectral window} & \colhead{Calibrators\tablenotemark{d}} & \colhead{Position} & \colhead{rms Cont./Line\tablenotemark{f}} \\[-3mm]
 & & \colhead{G358 (min.)} & \colhead{(mm)} & \colhead{(kilo-$\lambda$)} & \colhead{ frequencies\tablenotemark{c}(GHz)} & \colhead{} & \colhead{offset\tablenotemark{e}} &\colhead{(\mjb\/)}
 }
\startdata
\cutinhead{SMA}
240/345 & 2019-03-14 & 187 & 0.3 & 24--344 & 202.0 (A), 218.0 (A) & J1700-261/J1733-130 (Ga,Fl), 
& $-0\farcs144$, $-0\farcs012$ & 0.48 / 120\\
 & & & & & 290.3 (A), 306.3 (A) & J1744-312, 3C279 &  \\
345/400 & 2019-03-22  & 134 & 0.4 & 18--200 & 334.0 (A), 342.0 (A) & J1700-261/J1733-130 (Ga,Fl), & $-0\farcs075$, $-0\farcs088$ & 2.0 / 300 \\
 & & & & & 350.0 (A), 358.0 (A) & 3C279, J1924-292 &  \\
\cutinhead{ALMA}
Band 5 & 2019-04-16  & 55 & 0.9 & 9--494 & 189.033 (B), 189.523 (B) & J1924-292 (Fl), J1744-312 (Ga) & $0\farcs0$, $0\farcs0$ & 0.14 / 2.8 \\  
                                & &  & & & 199.610 (C), 200.925 (C), & & \\
                                & &  & & & 201.765 (D), 202.295 (D)  &  & \\
Band 6 & 2019-04-16  &  55 & 0.9 & 11--590 & 225.936 (E), 229.629 (F) & J1924-292 (Fl), J1744-312 (Ga) & $+0\farcs015$, $+0\farcs011$ & 0.12 / 2.5\\
                                   & & & & & 240.762 (E), 241.622 (E) &  & \\
Band 7 & 2019-04-12  &  49 & 0.6 & 16--725 & 330.231 (G), 330.858 (G),  & J1924-292 (Fl), J1744-312 (Ga) & $-0\farcs007$, $-0\farcs033$ & 0.35 / 7.2\\
                                  & & & & &  331.814 (H), 342.943 (H), & & \\
                                  & & & & &  343.660 (G), 344.373 (G) &  & \\
\enddata
\vspace{-2mm}
\tablenotetext{a}{Mean precipitable water vapor at zenith}
\vspace{-3mm}
\tablenotetext{b}{Range of projected baseline lengths} 
\vspace{-3mm}
\tablenotetext{c}{The bandwidth (MHz), and channel width (kHz) are indicated by the following letter codes:
A: 8000, 140; B: 117, 61; C: 234, 122; D: 468, 244;
E: 937.5, 244; F: 234, 61;
G: 117, 122 ; H: 937.5, 488}
\vspace{-3mm}
\tablenotetext{d}{Absolute flux and complex gain calibrators are marked with (Fl) and/or (Ga), respectively. All listed calibrators were used for SMA bandpass calibration, while J1924-292 was used for the ALMA observations.}
\vspace{-3mm}
\tablenotetext{e}{The Band 5 ALMA continuum image was used as the reference to remove small position offsets from the other datasets; these offsets were also applied to the line data.}
\vspace{-3mm}
\tablenotetext{f}{Aggregate continuum rms noise and representative noise per 0.21 or 0.12 \kms\/ channel for SMA and ALMA, respectively.}
%
\end{deluxetable*}

The recent discoveries of powerful accretion outbursts in two high-mass protostars -- NGC6334I-MM1
\citep{Hunter17,Hunter18} and S255IR-NIRS3 \citep{Caratti17,Liu18} -- have provided crucial insights on the role of
episodic accretion in massive star formation  \citep{Brogan18,Cesaroni18}.  A recent study of Orion suggests that
episodic accretion accounts for $\gtrsim$25\% of a star's mass \citep{Fischer19}, suggesting it is an important process
in star formation.  Both aforementioned outbursts were heralded by 6.7~GHz \methanol\/ maser flares
\citep{MacLeod18,Szymczak18,Fujisawa15},  and subsequent high resolution studies confirmed that the
flares occurred in dense gas surrounding the outbursting protostar \citep{Hunter18,Moscadelli17}, 
motivating new hydrodynamic simulations of protostellar accretion \citep{Meyer19a,Meyer19b}. 
Because Class~II \methanol\/ masers are pumped by infrared radiation \citep{Cragg05}, their apparent association with protostellar luminosity outbursts has a theoretical basis.  Thus, identifying and characterizing more events is critical to understanding massive star formation, which has inspired the international maser community to form the M2O group\footnote{See the Maser Monitoring Organization (M2O) website at {\tt MaserMonitoring.org}} to coordinate single-dish monitoring of masers and to perform rapid interferometric follow-up \citep[e.g.,][]{Burns18}.  
As a result, we are now better positioned to identify new outbursts, and catch them earlier in their evolution.  

On 2019 January 14, the 6.7~GHz Class~II \methanol\/ maser line in the massive star-forming region \gtfe\/ began flaring, rising in flux by an order of magnitude to 99\,Jy after two weeks \citep{Sugiyama19}. Maser emission in this transition toward this region was previously reported by \citet[][epoch 2006.24]{Caswell10} with a peak flux density of 10~Jy at \vlsr=$-15.9$~\kms\/ 
and by \citet[][epoch 2015.69; 1.7~Jy at $-18.7$~\kms]{Rickert19}.
This region appears as a compact clump in single dish (sub)millimeter continuum surveys including the 1.1~mm BOLOCAM Galactic Plane Survey (BGPS) \citep[G358.936-00.032,][]{Rosolowsky10} and the 0.87\,mm ATLASGAL survey \citep[AGAL358.931-00.029,][]{Urquhart13}. Otherwise, this field has been poorly studied, with no prior (sub)millimeter interferometric observations. 



%


Since the 6.7~GHz \methanol\/ maser flare in \gtfe\/ may be indicative of a massive protostellar outburst event, the M2O group has pursued multi-wavelength follow-up with telescopes worldwide, leading to the unprecedented discovery of several never-before-seen Class\,II \methanol\/ maser lines, include some in the torsionally-excited (v$_t$=1) state.  The first published results are the centimeter wavelength lines \citep{Breen19,Volvach19}. Light curves of the more common (v$_t$=0) maser lines from Hartebeesthoek Radio Observatory \citep{MacLeod19} show that the emission rose more rapidly than in NGC6334I-MM1 \citep[8~months,][]{MacLeod18} and S255IR-NIRS3 \citep[5~months,][]{Szymczak18}. While some velocity components have peaked, they persisted longer than the initial rise time, and other velocity components have emerged.  Thus, in contrast to the earlier cases, we have caught this flare during its initial rise, and additional follow-up is ongoing \citep[Bayandina et al, in prep.; Burns et al., in prep.,][]{Chen2019}. In this paper, we report first results from 0.87 to 1.5~mm  Submillimeter Array (SMA) and Atacama Large Millimeter/submillimeter Array (ALMA) observations taken $\sim +2$ and $\sim +3$~months after the 6.7\,GHz maser flare began. 


\begin{figure*}[th!]
\begin{minipage}{0.49\linewidth}
\includegraphics[width=0.99\linewidth]{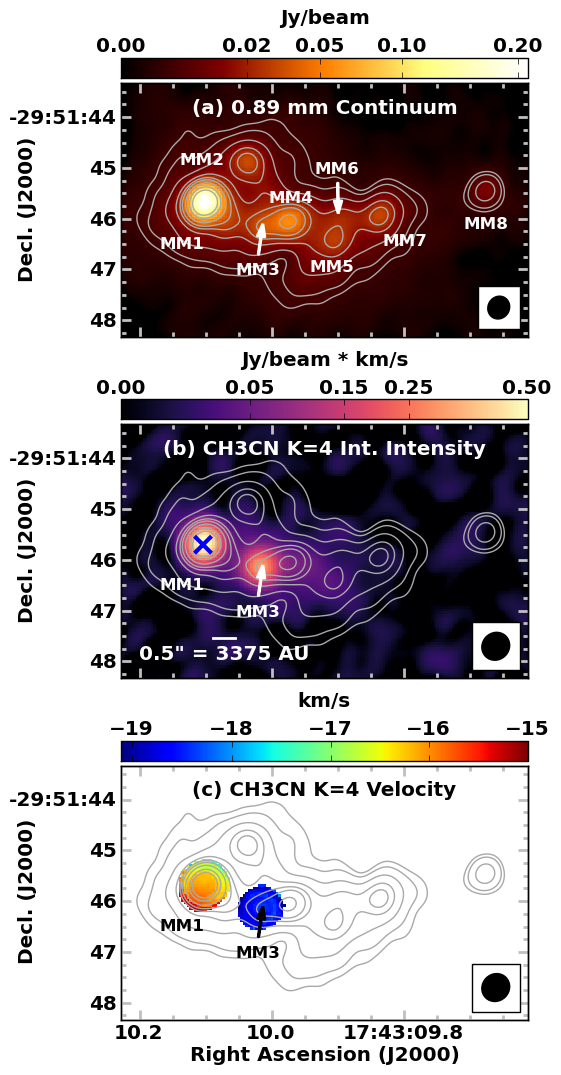}
\end{minipage}
\begin{minipage}{0.5\linewidth}
\includegraphics[width=0.535\linewidth]{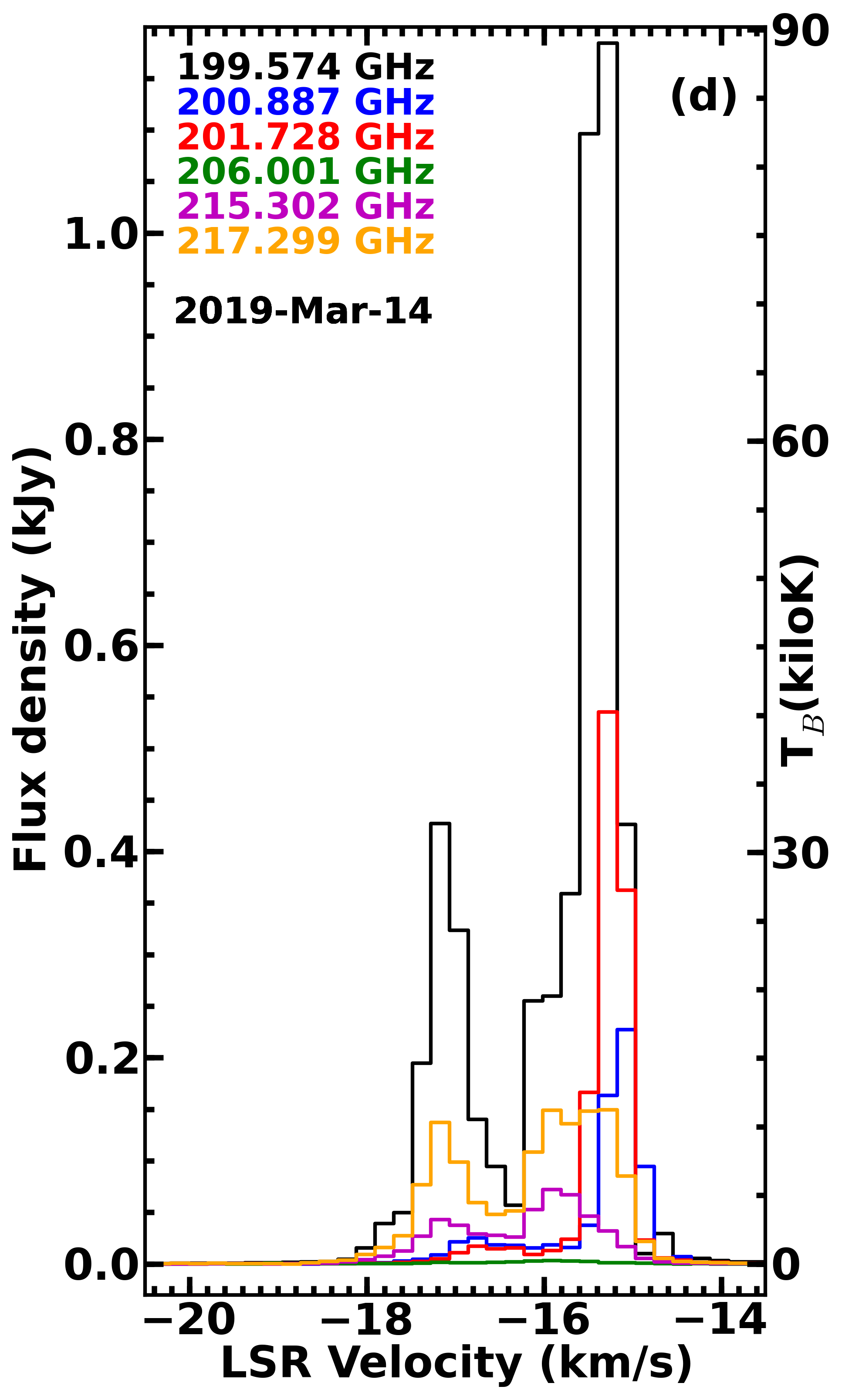} 
\includegraphics[width=0.455\linewidth]{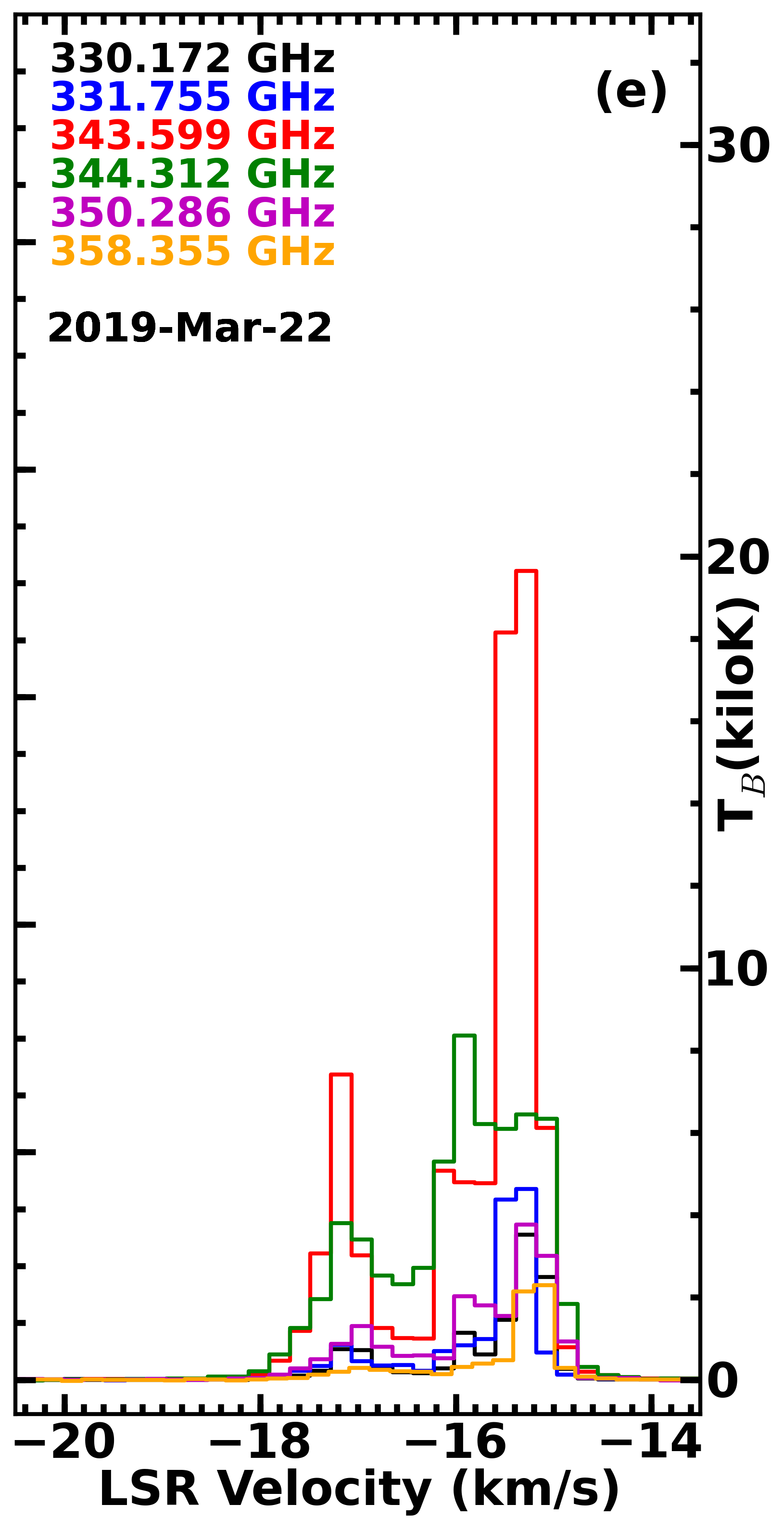} 
\includegraphics[width=0.545\linewidth]{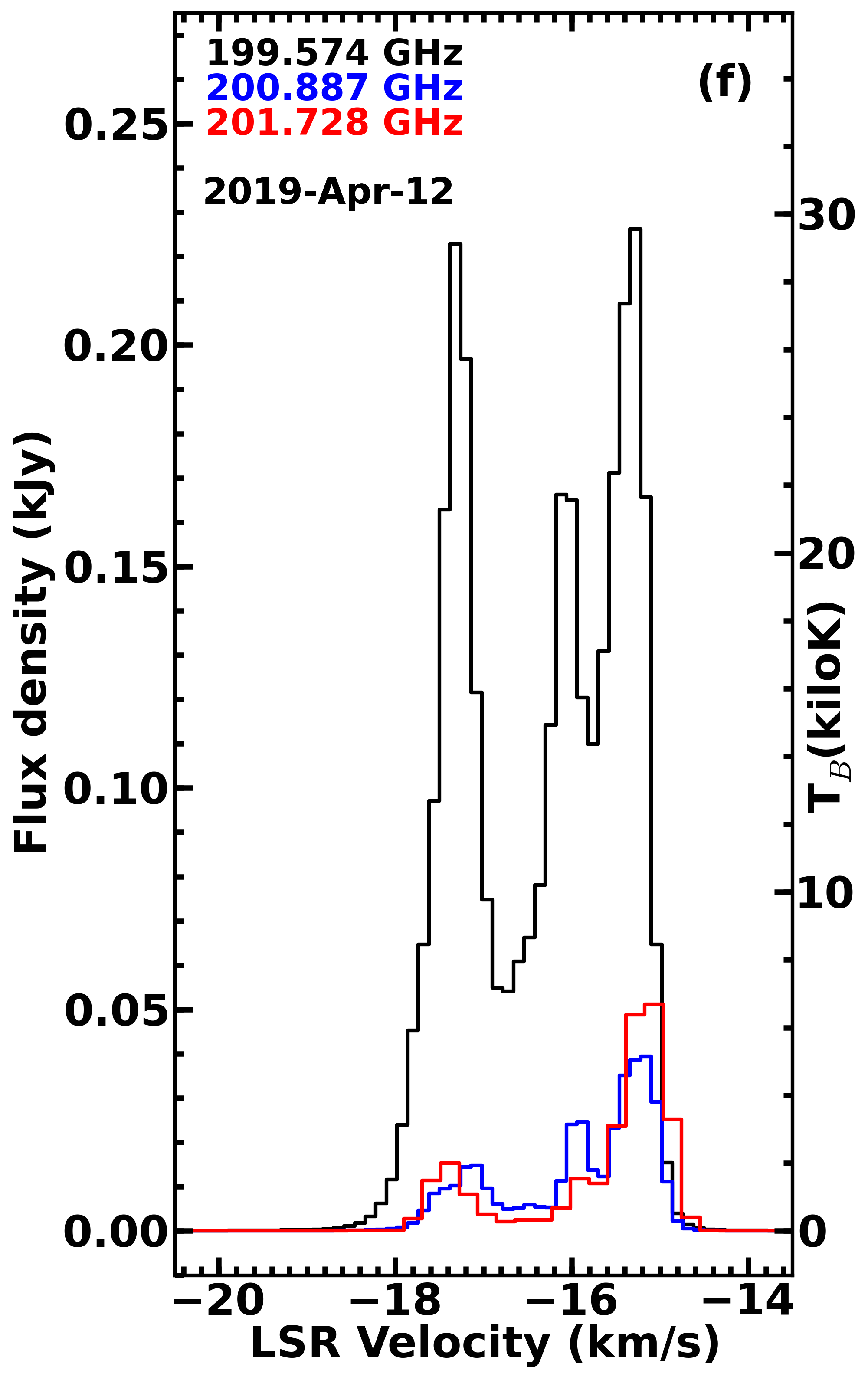} 
\includegraphics[width=0.445\linewidth]{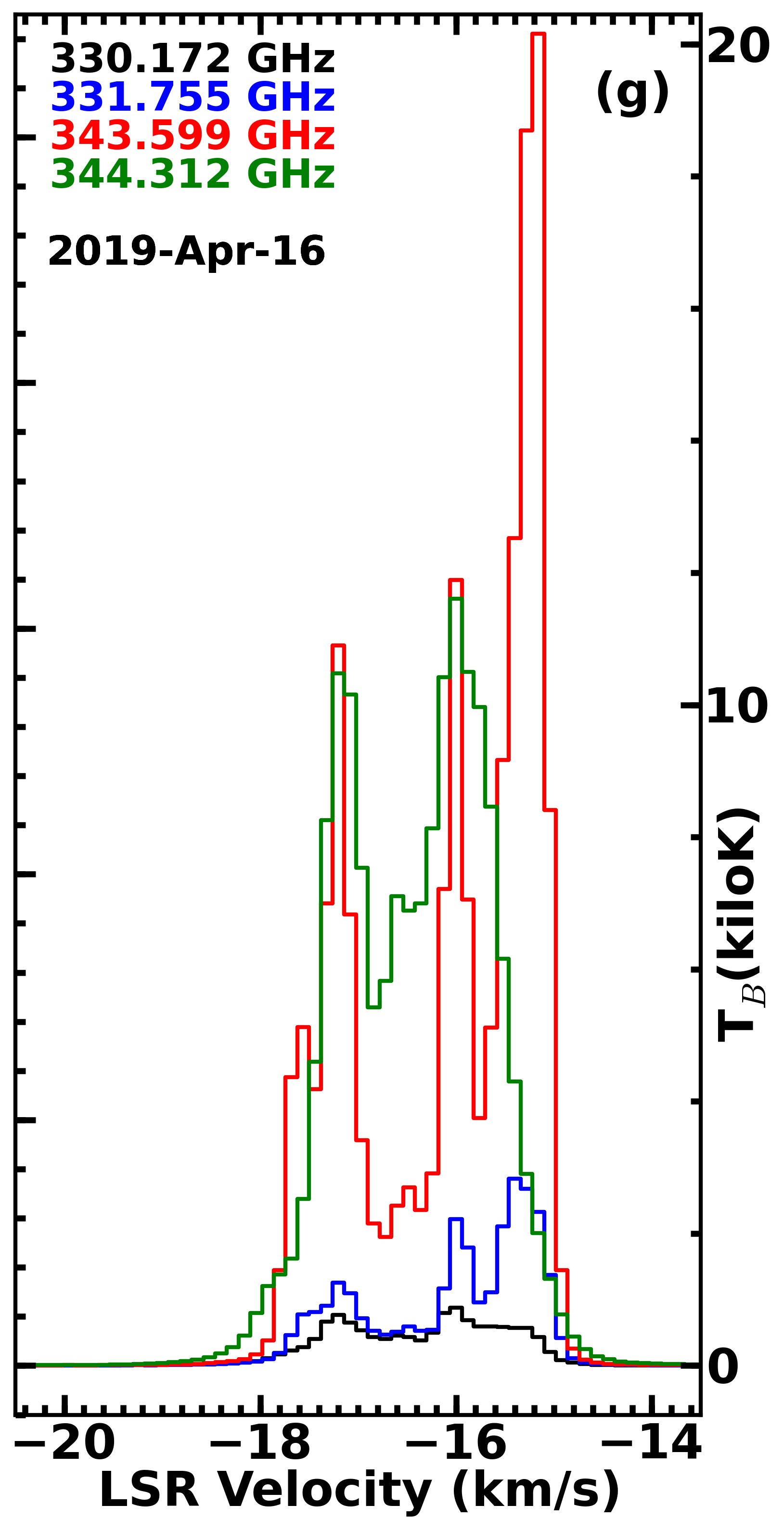} 
\end{minipage}
\caption{(a) ALMA 0.89\,mm (Band 7) continuum image of \gtfe, with an angular resolution $0\farcs46\times 0\farcs42$ ($\sim 3000$~au at 6.75\,kpc). Contour levels are 3.5\,\mjb\/ ($1\sigma$)$\times(8,12,24,48,72,96,144,240,336)$; the  continuum components are labeled for reference. Integrated intensity (b) and velocity (c) images of the CH$_3$CN $J=11_4-10_4$ transition (Band 5 data, beam $0\farcs55$), with the 0.89\,mm contours from (a) overlaid. The location of the (sub)millimeter masers is indicated by the blue $\times$ symbol in (b). 
(d-g) Spatially integrated maser spectra for (d) SMA, near 200~GHz; (e) SMA, near 340~GHz; (f) ALMA, for three of the same lines as panel d; (g) ALMA, for four of the same lines as panel e. The brightness temperature (T$_b$) scales (shown in kiloKelvin) are computed with respect to the synthesized beam, not the fitted size.}
\label{spectra}
\end{figure*}

\begin{deluxetable*}{cccccccccc}
\tabletypesize{\scriptsize}
\tablecolumns{10}
\setlength{\tabcolsep}{0.5mm}
\tablecaption{Continuum Source Properties \label{tab:fluxes}}
\tablehead{
\colhead{Source}  & \multicolumn{2}{c}{J2000 Fitted Position\tablenotemark{a}} & \colhead{Fitted Size\tablenotemark{b}} & \multicolumn{3}{c}{ALMA Flux Density} & \colhead{Spectral} & \colhead{SMA Flux} & \colhead{Percent\tablenotemark{c}} \\[-3mm]
& \colhead{R.A.} & \colhead{Dec.}   & \colhead{major$\times$minor,\,pa ($\sigma_{size}, \sigma_{pa}$)} & \multicolumn{3}{c}{(mJy)} & \colhead{Index} & \colhead{Density (mJy)} & \colhead{Diff.} \\[-3mm]
           &  &     & \colhead{$''\times '', \arcdeg$ ($'', \arcdeg$)} & \colhead{195.58 GHz} & \colhead{233.75 GHz} & \colhead{337.26 GHz} &  Fit & \colhead{210.88 GHz} &  \colhead{Epochs} 
}
\startdata
MM1 & 17:43:10.1014 (0.0005) & -29:51:45.693 (0.002)  &  $0.26\times 0.22, +80$ ($0.02, 2$) &  52.5 (0.3) &  96.7 (0.3) &  281.8 (0.8) &  3.06 (0.13) &  70.1 (0.9) & $+3.3$   \\
MM2 & 17:43:10.036 (0.001) &   -29:51:44.935 (0.005)  &  $0.51\times 0.41, +67$ ($0.06, 4$) &   8.7 (0.4) &  18.3 (0.4) &   64.4 (1.2) &  3.60 (0.15) &  11.0 (1.3) & $-8.4$ \\
MM3 & 17:43:10.024 (0.002) &   -29:51:46.123 (0.02)   &  $0.56\times 0.32, +50$ ($0.04, 13$) &  5.4 (0.4) &  10.4 (0.4) &   39.9 (1.1) &  3.67 (0.17) &   7.1 (0.9) & $+0.0$    \\
MM4 & 17:43:09.975 (0.001) &   -29:51:46.065 (0.009)  &  $0.55\times 0.29, +82$ ($0.04, 2$) &  15.4 (0.4) &  26.0 (0.4) &   93.5 (1.1) &  3.34 (0.13) &  20.7 (1.3) & $+7.4$ \\  
MM5 & 17:43:09.908 (0.001) &   -29:51:46.435 (0.005)  &  $0.57\times 0.42, +137$ ($0.03, 9$) &  7.9 (0.4) &  13.4 (0.4) &   50.6 (1.2) &  3.47 (0.16) &  10.0 (1.5) & $+1.7$    \\
MM6 & 17:43:09.900 (0.008) &   -29:51:45.92 (0.07)    &     $< 0.4$                                    &   1.6 (0.2) &   3.1 (0.2) &    9.9 (0.4) &  3.28 (0.22) &   2.6 (0.4) & ...   \\
MM7 & 17:43:09.841 (0.001) &   -29:51:45.96 (0.05)    &  $0.52\times 0.35, +122$ ($0.07, 2$) &  8.9 (0.4) &  14.6 (0.3) &   62.5 (1.2) &  3.68 (0.15) &  11.1 (1.3) & $+2.1$    \\
MM8 & 17:43:09.677 (0.001) &   -29:51:45.49 (0.05)    &    $<0.2$                                       &  2.9 (0.3) &   4.9 (0.2) &   13.3 (0.7) &  2.77 (0.21) &   3.4 (0.9) & ...    \\
\enddata
\vspace{-2mm}
\tablenotetext{a}{From the 195.58\,GHz ALMA Band 5 fits, the $1\sigma$ uncertainty includes the dispersion among the ALMA bands.}
\vspace{-3mm}
\tablenotetext{b}{Error-weighted fitted mean deconvolved size from the three ALMA bands; the $1\sigma$ uncertainties include the dispersion between bands and fitting uncertainty added in quadrature. An upper limit is given when the deconvolved size could not be measured.}
\vspace{-3mm}
\tablenotetext{c}{Percent difference of the SMA 210.88 GHz flux density (epoch 2019.20) relative to the value predicted at this frequency by the ALMA spectral index (epoch 2019.29); sources with SMA flux measurements $<7\sigma$ are excluded.}
\end{deluxetable*}



\section{Observations} 


The observing parameters of our
SMA and ALMA data are summarized in Table~\ref{tab:observations}. 
The SMA data were calibrated in MIR, and the ALMA data were calibrated using the ALMA CASA pipeline. The SMA employed 8 antennas while ALMA employed 43, and the calibrators in each observation are listed in Table~\ref{tab:observations}. For consistency, ALMA flux monitoring data was used to set the absolute flux scale for all the observations\footnote{ALMA Source Catalog:  \url{https://almascience.eso.org/sc}}. We estimate that the absolute flux calibration uncertainty is $5\%$ and $10\%$ for the ALMA and SMA data, respectively. All of the data were imaged in CASA 5.4.0-70, and self-calibrated using the strongest maser channel in each respective band. The SMA 290.3 and 306.3\,GHz windows did not contain any maser emission and are not discussed further.

The SMA continuum images were made with robust weighting $R=+0.5$ and have beam sizes of $0\farcs66\times 0\farcs46$ (pa$=+4\arcdeg$) at 210.88\,GHz and $1\farcs1\times 0\farcs7$ (pa$=+14\arcdeg$) at 346\,GHz.
The ALMA continuum images were made with $R=-0.5$, $+0.1$, and $+0.1$ for Bands 5, 6, and 7 (1.5, 1.3, and 0.89\,mm), respectively, yielding similar angular resolutions that were subsequently convolved to a common beam of $0\farcs46\times 0\farcs42$ (pa$=-30\arcdeg$) to measure the individual source properties. An $R=+1.0$ Band 7 image was also created to measure the integrated flux density for comparison to single dish measurements. Small systematic position offsets among the continuum images were removed using the ALMA Band 5 195~GHz image as the reference (see Table~\ref{tab:observations}). The ALMA Band 5 absolute position uncertainty is estimated to be $<40$~mas.

\begin{deluxetable*}{cccccccccc}[ht!]
\tablecolumns{10}
\tabletypesize{\footnotesize}
\setlength{\tabcolsep}{0.1cm}
\tablecaption{Properties of detected \methanol\/ maser lines\tablenotemark{a} \label{tab:masers}}
\tablehead{
\colhead{Torsional} & \colhead{Transition} & \colhead{Rest} & \colhead{\eupper\tablenotemark{b}} & \multicolumn{2}{c}{Channel spacing} & \colhead{SMA (2019 Mar)} & \multicolumn{2}{c}{ALMA (2019 Apr)} & 
\colhead{Integrated Flux} \\[-3mm]
\colhead{State} & \colhead{Quantum} & \colhead{Frequency} & & \colhead{SMA} & \colhead{ALMA} & \colhead{Integrated Flux} & \colhead{Integrated Flux} & \colhead{T$_b$\tablenotemark{c}} & \colhead{Ratio}\\[-3mm]
\colhead{\vt} & \colhead{Numbers} & \colhead{(GHz)} & \colhead{(K)} & \colhead{(\kms)} & \colhead{(\kms)} &  \colhead{(Jy~\kms)} & \colhead{(Jy~\kms)} & \colhead{(K)} & \colhead{SMA/ALMA}
}
\startdata
1 & $13_{-2}-14_{-3}~E2$   & 199.574851(18) & 575.2   & 0.21 & 0.12 &  980  & 314.0 &  3.17E+6 & 3.12 \\ 
1 & $18_3-19_4~E1$         & 200.887863(31) & 812.5   & 0.21 & 0.12 &  137  & 40.5  &  1.67E+6 & 3.38 \\ 
1 & $16_{-1}-17_{-2}~E2$   & 201.728147(29) & 728.2   & 0.21 & 0.21 &  250  & 44.9  & 1.18E+6  & 5.56 \\ 
0 & $12_5-13_4~E1$         & 206.001302(15) & 317.1   & 0.21 & ...  &  5.0  &  ...  & ...      & ... \\  
1 & $6_1-7_2~A^+$   & 215.302206(19) & 373.8   & 0.21 & ...  &  94.3   &  ...  & ...      & ... \\  
1 & $6_1-7_2~A^-$   & 217.299205(17) & 373.9   & 0.21 & ...  & 246   &  ...  & ...      & ...\\   
0 & $15_4-16_3~E1$         & 229.589056(12) & 374.4   & ...  & 0.12 &  ...  & 8.6   &  4.82E+3    & ... \\
2\tablenotemark{d} & $11_3-12_4~A^-$ & 330.172526(22) & 810.7   & 0.21 & 0.12 & \multirow{2}{*}{\big\} 118}  & \multirow{2}{*}{16.4}  &  \multirow{2}{*}{1.39E+4}   & \multirow{2}{*}{7.20} \\ 
2\tablenotemark{d} & $11_3-12_4~A^+$ & 330.172553(22) & 810.7   & 0.21 & 0.12 &   &  &    & \\ 
1 & $15_{-5}-16_{-6}~E2$   & 331.755099(32) & 823.9   & 0.21 & 0.12 &  147  & 37.2  & 1.22E+5   & 3.95 \\ 
1 & $13_{-1}-14_{-2}~E2$   & 343.599019(26) & 624.0   & 0.21 & 0.12 &  738  & 204.7 & 1.35E+6  & 3.61 \\ 
1 & $10_{-2}-11_{-3}~E2$   & 344.312267(17) & 491.9   & 0.21 & 0.12 &  559  & 187.0 &  9.27E+4   & 2.99 \\ 
1 & $15_{3}-16_{4}~E1$     & 350.286493(25) & 694.8   & 0.21 & ...  &  174  & ...   & ...      & ...  \\
1\tablenotemark{d} & $18_4-19_5~A^-$ & 358.354940(27) & 877.2 & 0.21 & ...  & \multirow{2}{*}{\big\} 68.0} & ... & ... & ...\\
1\tablenotemark{d} & $18_4-19_5~A^+$ & 358.355121(27) & 877.2 & 0.21 & ...   &    & ... & ...   & ...\\ 
1 & $3_1-4_2~A^-$   & 361.236506(17) & 339.2   & 0.21 & ...  &   11.1  & ...   & ...      & ...    
\enddata
\vspace{-2mm}
\tablenotetext{a}{Transition parameters were extracted from the JPL Line Catalog \citep{Pickett98} compiled from \citet{Xu08}.}
\vspace{-3mm}
\tablenotetext{b}{Energies are relative to the lowest rotational level ($0_0$) of v$_t$=0 $A$-type methanol at 0 K.}
\vspace{-3mm}
\tablenotetext{c}{Peak Rayleigh-Jeans brightness temperature computed from peak intensity and deconvolved size (typically  0.1--0.3 of the beam). }
\vspace{-2mm} 
\tablenotetext{d}{Blended}
\end{deluxetable*}

\begin{figure*}[h!]   
\includegraphics[width=0.99\linewidth]{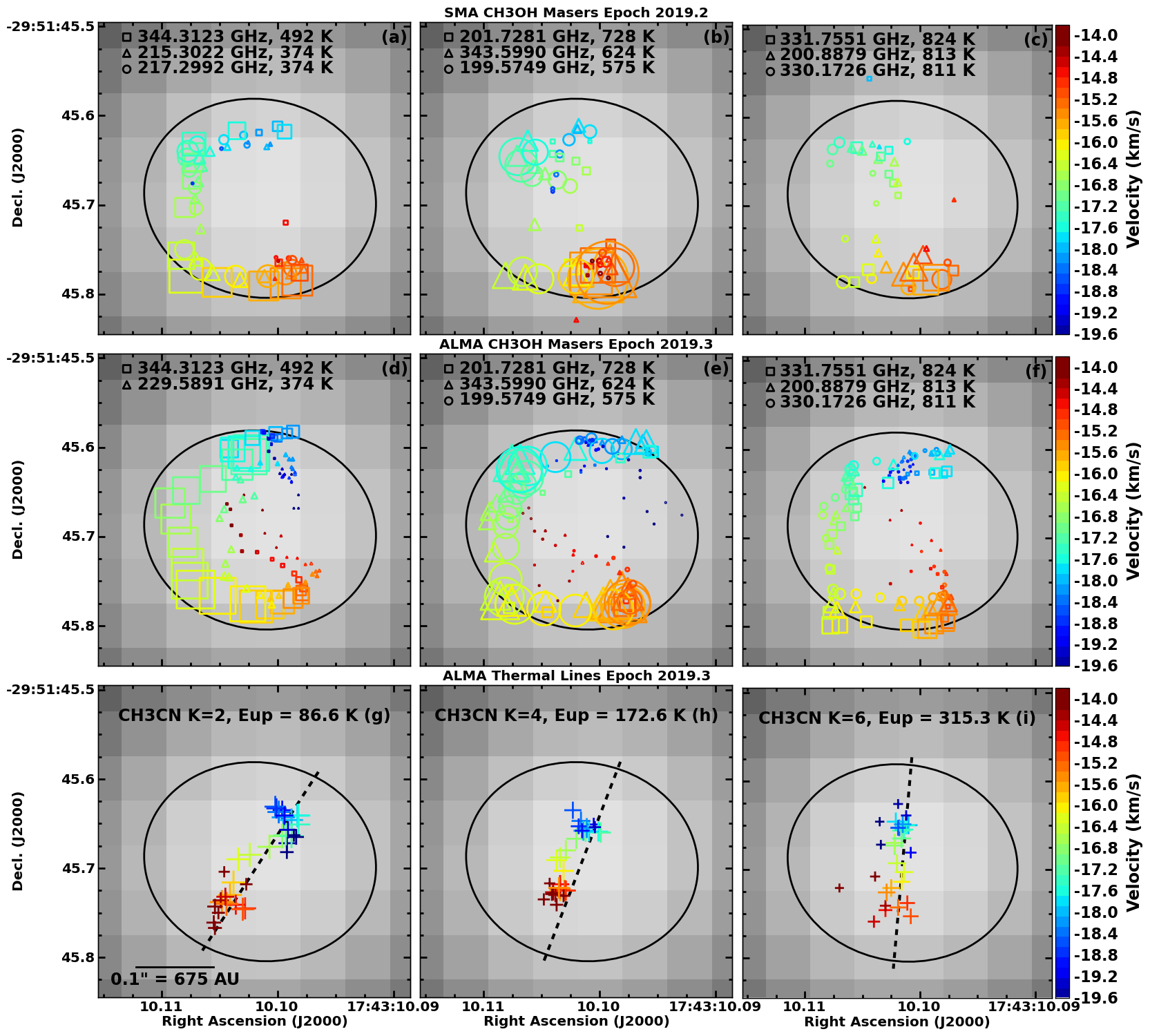}
\caption{In each panel, the spatio-kinematics of the indicated line(s) is (are) shown on the 0.89\,mm continuum image from Fig~\ref{spectra}a  (pixel size 50~mas), zoomed in towards MM1. Each plotted symbol shows the location of the peak emission from a 2D Gaussian fit in each channel ($\leq 0.21$\,\kms\/ wide; the symbols are color-coded by velocity.  The fitted relative position uncertainty is 20~mas, while the absolute uncertainty is twice that. Each panel also shows an ellipse representing the deconvolved size (FWHM) of the millimeter continuum emission from MM1  (Table~\ref{tab:fluxes}). Panels (a-c) show \methanol\/ maser lines from SMA, panels (d-f) show \methanol\/ maser lines from ALMA, and panels (g-i) show thermal CH$_3$CN $J=11-10$ lines from ALMA.  Symbol size is proportional to intensity on the same relative scale within each row; for SMA the maser symbol size is 1/3 the ALMA maser symbol size. Panels (g-i) also show position angle vectors (dashed) of $-33.0\arcdeg$,  $-21.0\arcdeg$, and $-5.0\arcdeg$, respectively. 
}
\label{spots}
\end{figure*}

\section{Results}

\subsection{(Sub)millimeter Continuum and Thermal Line Emission}
\label{continuum}
As shown in Figure~\ref{spectra}a, we resolve \gtfe\/ into a protocluster of eight (sub)millimeter continuum sources at an angular resolution of $0\farcs46\times 0\farcs42$
at 0.89~mm. These sources have been designated MM1...MM8 in order of decreasing right ascension. The morphology is consistent from 0.89 to 1.5~mm and across the SMA and ALMA epochs. The projected linear separations between nearest neighbors are a few thousand to 10,000 au (assuming a distance 6.75\,kpc, see below for details), which is typical of other regions of massive star formation \citep{Brogan16,Beuther18}.  The integrated flux density of \gtfe\/ at 0.89~mm from the $R=+1.0$ ALMA image is $1.13\pm0.03$~Jy. 

The continuum images from the three ALMA bands, and the SMA 210.88~GHz image, were fit with eight 2D Gaussian components (the resolution of the 346\,GHz SMA data is too poor to resolve the majority of sources); the fitted properties are provided in Table~\ref{tab:fluxes}. 
After subtracting the model components, two areas of residual emission are consistently present across all the images. The brightest is from diffuse emission between MM1 and MM3, which appears to be a ``bridge'' connecting the two and not a distinct source; the second is compact and located at the peak position of MM1, suggesting additional unresolved structure not well-fit by a single Gaussian. For the ALMA 1.5 to 0.89\,mm images, the ``bridge'' has peak intensities of 3.4, 7.0, and 23.5\,\mjb\/, while the MM1 residual has peak intensities of 2.4, 5.1, and 18.7\,\mjb\/.  

Fig.~\ref{spectra}b shows the integrated intensity of the CH$_3$CN $J=11_4-10_4$ transition (\eupper=172.5\,K), revealing that \gtfe\/ harbors two molecular hot cores coincident with MM1 and MM3, with MM1 showing significantly richer spectra.  Using hot core tracers, the center velocity of MM1 is \vlsr=$-16.5\pm0.3$ \kms\/ and the full width at half maximum line width is $\Delta V=3.1\pm 0.2$ \kms\/. For MM3, \vlsr=$-18.6\pm0.2$ \kms\/ and $\Delta V=3.7\pm 0.2$ \kms\/.  Fig.~\ref{spectra}c demonstrates that hot core tracers like CH$_3$CN show a roughly north-south velocity gradient across MM1, while MM3 shows little velocity variation at the current angular resolution. 

Lacking a maser parallax distance, we use the MM1 \vlsr\/ to derive a near kinematic distance of 6.75\,kpc \citep[model~A5,][see \url{http://bessel.vlbi-astrometry.org/revised_kd_2014}]{Reid14}. This estimate has considerable uncertainty ($+0.37, -0.68$\,kpc reported by the model) due to the proximity of the source to the Galactic center. 

The MM1 fitted continuum size ($0\farcs26\times0\farcs22$, Table~\ref{tab:fluxes}) is near the lower limit that can be probed with the current resolution data, thus the true source size may well be smaller. Indeed, the methyl cyanide (CH$_3$CN) $J=11-10$, $K=0-6$ emission is well-matched to a single excitation temperature model with $T_{ex}=172\pm 3$\,K and a background temperature $T_{bg}=159\pm 3$\,K \citep[see methodology in][]{McGuire18}. This model also reproduces the emission from other hot core tracing molecules like glycolaldehyde, ethylene glycol, methyl formate v=0 and 1, ethanol, and acetaldehyde (full analysis of the hot core line emission will be provided in a future work).   Given the observed MM1 peak brightness temperature $T_b=53$\,K in the 0.89\,mm ALMA image, and assuming that the dust continuum is becoming optically thick at 0.89\,mm (so that $T_b\approx$ physical temperature when the emission is resolved), this implies that the size of the dust emission region is $\sim 0\farcs14$ ($\sim 940$\,au). 

\subsection{(Sub)millimeter \methanol\/ Masers}

The 14 new Class~II \methanol\/ maser lines first detected in these data are presented in
Table~\ref{tab:masers} and Figs.\ref{spectra}(d-g). 
All of the newly discovered masers are located near the peak of MM1, the brightest continuum source (Fig.~\ref{spectra}b). The 
new maser transitions have \eupper=317.1 to 877.2~K; the majority are from rotational levels within the \vt=1 torsionally excited state. One maser arises from a \vt=2 transition, providing the first observational evidence that Class\,II maser pumping involves levels in the \vt=2 state \citep{Sobolev94}.
The brightest maser lines observed with ALMA have $T_b>10^6$\,K. For the seven maser lines that are in common between the SMA and ALMA, we find that the integrated flux density has dropped by a factor of 3 to 7 during the intervening month; though the velocity extent of maser emission stayed constant. The v$_t$=0 line at 349.1070\,GHz, identified as a maser in S255IR-NIRS3 \citep{Zinchenko17}, was covered in our SMA tuning but not detected.

Figure~\ref{spots} shows the spatio-kinematic (combined spatial morphology and kinematics) behavior of the SMA and ALMA \methanol\/ maser lines and representative thermal lines from the ALMA data. For each plotted line, a two-dimensional Gaussian was fitted to each channel with emission. Channels with signal-to-noise ($S/N$) $< 12$ and $< 16$ for ALMA and SMA, respectively, are omitted to ensure relative position uncertainties better than 20\,mas.  In both epochs, the (sub)millimeter \methanol\/ masers consistently trace a roughly elliptical curvilinear pattern centered on the continuum peak, with major and minor axes of $0\farcs20$ and $0\farcs17$ ($\approx$1 light-week), and position angle $+119\arcdeg$.  The brightest masers are located across from each other, to the northeast (blueshifted) and southwest (redshifted).  An additional set of spots extending northeast of the continuum peak appears only in the first epoch, only in the highest excitation lines, and primarily at the central \vlsr\/ (Fig.~\ref{spots}a,b).   In contrast, the thermal hot core-tracing lines reveal a different morphology, tracing a linear pattern $0\farcs15$  ($\sim 1000$\,au) in length that crosses the continuum peak. Interestingly, for CH$_3$CN $J=11-10$, the position angle of this feature changes from $\sim -33\arcdeg$ in the lowest (coldest) $K$=2 transition to $\sim -5\arcdeg$ in the highest (hottest) $K$=6 transition. This progression could result from a temperature and/or opacity effect, coupled with the emitting region's geometry. 

\subsection{\gtfe\/ Luminosity}

Assuming a distance of 6.75\,kpc (\S 3.1), along with photometry extracted from 1.1\,mm BGPS \citep{Rosolowsky10}, 0.87\,mm ATLASGAL \citep{Schuller09}, and 70-500\,\mum\/ {\em Herschel} HiGAL \citep{Molinari16} survey images for the integrated emission from the \gtfe\/ region, a single temperature greybody fit yields $T_{dust}=28.5\pm 1.5$\,K, implying an L$_{\rm FIR} \approx 7660$~\lsun, following the procedure of \citet{Towner19}. We also fit the full SED, including the  MIPSGAL 24\,$\mu$m detection \citep{Gutermuth15}, to the model grid of young stellar objects of \citet{Robitaille17}, using the method described in \citet{Towner19}. The best fits yield L$_{\rm bol}$ of $\approx20000$~\lsun, 
but because the shorter wavelength data may be contaminated with foreground emission, this value should be considered an upper limit. Accounting for the distance uncertainty, the plausible luminosity range (for the near distance) is 5700-22000~\lsun.  We measured an 0.87\,mm ATLASGAL flux density of $1.10\pm 0.16$\,Jy, after making a correction for an estimated $17\%$ line contamination based on the integrated (non-maser) line emission in the ALMA Band 7 data.  With this flux density and the $T_{dust}$ estimate, we find a total gas mass of $167\pm 12$\,M$_{\odot}$ for \gtfe\/ protocluster (assuming a grain opacity spectral index $\beta=1.7$ and dust-to-gas ratio of 100).  

\section{Discussion}



Maser emission will arise from locations where the physical conditions are favorable for the maser pumping mechanism, and the line-of-sight affords a velocity-coherent high column density path length. The rapid evolution of the maser flare in \gtfe\/, with dramatic changes over the course of only 1 month, together with the large spatial scale over which changes have occurred ($\sim 1200$\,au), require that the impetus for change must be radiative: the timescale for physical movement or bulk changes in the line-of-sight path is far too long. For example, at 6.75~kpc a parcel of gas moving at 100 \kms\/ would take 16 years to traverse 50 mas ($\sim 340$\,au, 1 pixel of Fig.~\ref{spots}).

One explanation for the maser flare is an abrupt change to more favorable physical conditions, such as the rapid heating of dust and gas that occurs in a protostellar accretion event \citep{Johnstone13}. Comparing our measurement of the 0.87\,mm ATLASGAL flux density (1.11\,Jy) with the total ALMA 0.89\,mm measurement after scaling to 0.87\,mm, 1.22$\pm$0.04~Jy (using the flux-weighted mean dust spectral index of +3.3), yields a post-flare excess of 0.11$\pm$0.19~Jy at 0.87\,mm, which is consistent with no change.  Previous estimates of the ATLASGAL flux density for this source range from 1.09-1.4\,Jy \citep{Csengeri14,Contreras13}, with the former being the result of applying a spatial filtering technique, but neither correcting for line contamination. It is surprising that the spatially filtered flux density is reduced from the unfiltered values by $22-30\%$ given that this source is unresolved by ATLASGAL (the fitted size is smaller than the ATLASGAL beam). Nevertheless, using the spatially filtered value as a lower limit to the pre-flare flux density, and correcting for line contamination (\S~\ref{continuum}), yields 0.90$\pm$0.16~Jy, suggesting a maximum possible excess of 0.32$\pm$0.15~Jy. However, this $2\sigma$ result is larger than the current total flux density of MM1 scaled to 0.87\,mm (0.30$\pm$0.01~Jy), suggesting any change was considerably smaller.   

It is also feasible that a modest brightening of MM1, which currently represents 25\% of the total ALMA flux, is being masked by a commensurate loss in the total flux of the protocluster due to spatial filtering of larger-scale emission by ALMA compared to ATLASGAL's $19\farcs2$  beam. Unfortunately, the lack of prior higher resolution millimeter data prevents us from testing this possibility. The two post-flare (sub)millimeter observations do not show evidence for any significant ($>10\%$) change in the continuum flux of MM1 between the 2019.2 and 2019.3 epochs (Table~\ref{tab:fluxes}). In summary, the uncertainty in the pre- and post-flare flux densities (and spatial filtering) allow for a modest brightening to have occurred, but it remained steady between 2 and 3 months after the maser flare began, in contrast to the masers. Alternatively, because the heating and cooling timescale of the dust is shorter than the gas \citep{Johnstone13}, it is possible that modest dust heating was associated with the maser flare, but had already 
subsided before our SMA and ALMA observations. In contrast, the (sub)millimeter emission of S255IR-NIRS3 increased by a factor of 2 and subsided after 2 years, while \ngci\/ increased by a factor of 4 and has yet to subside after 4 years \citep[][Hunter et al.\ in prep.]{Liu19,Hunter17}

%


The spatio-kinematic pattern of a partial elliptical ring formed by the \methanol\/ masers, especially in epoch 2019.3 (Fig.~\ref{spots}d-f), suggests a coherent physical structure that has been illuminated by a radiative event from the protostar.  The pattern is not consistent with simple rotation of an inclined circular disk -- in that case the most red- and blue-shifted emission should be observed at the intersection of the ellipse with its major axis. The distribution of maser spots in \gtfe\/ is similar to ring-like structures in the 6.7\,GHz maser line originally found by \citet{Bartkiewicz05} toward the massive protostar G23.657$-$0.127, for which recent proper motion measurements suggest slow expansion rather than disk rotation \citep{Bartkiewicz18}. Such a morphology could arise from the walls of an outflow cavity. 

Interpreting the velocity gradient in the thermal CH$_3$CN emission as arising from disk rotation, as seen in other massive protostars \citep[e.g.][]{Ilee18}, also presents concerns. While the radius and linewidth would imply a dynamical mass of (5/{sin\,$i$})\,\msun, the higher velocity channels do not peak closer to the center of the distribution but instead near the edges, which suggests a ring morphology with edge-on inclination ($i\sim90\arcdeg$) for the thermal gas; or simply two sources at different $v_{lsr}$.  Interestingly, the faint redshifted spots in the ALMA maser spot diagrams (Fig.~\ref{spots}d-f) appear to connect with the redshifted side of the thermal gas velocity gradient.


One explanation for the apparent connection of the redshifted maser and thermal gas is a spiral filament of infalling gas, a concept invoked to explain the kinematic structures in thermal species (including \methanol\/) toward G10.6$-$0.04 \citep{Liu17}, and in the methanol masers surrounding Cepheus~A~HW2 \citep{Sanna17}.  Similarly, hydrodynamic simulations predict spiral shapes in fragmenting accretion disks around massive protostars, leading to large accretion outbursts as fragments reach the protostar, a mechanism proposed to explain the outburst in S255IR-NIRS3 \citep{Meyer18}.  Spiral arm structures can also be driven by an encounter between a protostellar disk and a massive companion \citep{Cuello19,Clarke93}, including cases that generate a single tidal arm.  

In conclusion, while it is clear that a significant radiative event must have occurred in \gtfe\/ to produce a strong but rapidly declining \methanol\/ maser flare with a coherent velocity pattern, coupled with the discovery of numerous new \methanol\/ maser lines, the lack of conclusive evidence for dust heating is perplexing.  
For certain, this event has different characteristics compared to the prior two massive protostellar outburst events (S255IR-NIRS3, NGC6334I-MM1). Ongoing monitoring of the maser decline \citep[also see][]{MacLeod19}, future observations of outflow tracers, higher angular resolution dust continuum observations, and detailed investigation of the maser pump conditions will provide essential clues to unravel the nature of this enigmatic maser flare event.


\acknowledgments

The National Radio Astronomy Observatory is a facility of the National Science Foundation operated under agreement by the Associated Universities, Inc. This paper makes use of the following ALMA data: ADS/JAO.ALMA\#2018.A.00031.T. ALMA is a partnership of ESO (representing its member states), NSF (USA) and NINS (Japan), together with NRC (Canada) and NSC and ASIAA (Taiwan) and KASI (Republic of Korea), in cooperation with the Republic of Chile. The Joint ALMA Observatory is operated by ESO, AUI/NRAO and NAOJ.    
The SMA is a joint project between the Smithsonian Astrophysical Observatory and the Academia Sinica Institute of Astronomy and Astrophysics and is funded by the Smithsonian Institution and the Academia Sinica.
This research used the \url{https://www.splatalogue.net} spectroscopy database \citep{Remijan07}.
\facility{ALMA, SMA}
\software{CASA \citep{McMullin07}, 
APLpy \citep{aplpy2012}}

\bibliography{bibliography}

\end{document}